\documentclass[multphys,vecphys]{svmult}

\usepackage{makeidx}         
\usepackage{graphicx}        
\usepackage{multicol}        
\usepackage[bottom]{footmisc}

\makeindex             
\newcommand{\oversim}[2]{\protect{\mbox{\lower0.5ex\vbox{%
  \baselineskip=0pt\lineskip=0.2ex
  \ialign{$\mathsurround=0pt #1\hfil##\hfil$\crcr#2\crcr\sim\crcr}}}}}
\newcommand{\simgreat}{\mbox{$\,\mathrel{\mathpalette\oversim>}\,$}} 
\newcommand{\simless} {\mbox{$\,\mathrel{\mathpalette\oversim<}\,$}} 
\begin{document}

\title*{The dynamical evolution of young clusters and galactic implications}
\author{Pavel Kroupa}
\institute{Argelander Institute for Astronomy, Auf dem H\"ugel 71,
D 53121 Bonn, Germany
\texttt{pavel@astro.uni-bonn.de}
}
%
%
\maketitle\footnote{to appear in the proceedings of 
"Globular Clusters: Guides to Galaxies", Concepcion, Chile, 
March 6th-10th, 2006, eds Tom Richtler et al. (Springer)}

Star clusters are observed to form in a highly compact state and with
low star-formation efficiencies. If the residual gas is expelled on a
dynamical time the clusters disrupt thereby (i) feeding a hot
kinematical stellar component into their host-galaxy's field
population, and (ii) if the gas-evacuation time-scale depends on
cluster mass, then a power-law embedded-cluster mass function
transforms within ten to a few dozen Myr to a mass function with a
turnover near $10^5\,M_\odot$, thereby possibly explaining this
universal empirical feature.

\section{Early cluster evolution}
\label{pk_sec:earlyevol}
The star-formation efficiency (sfe), $\epsilon\equiv M_{\rm
ecl}/(M_{\rm ecl} + M_{\rm gas})$, where $M_{\rm ecl}, M_{\rm gas}$
are the mass in freshly formed stars and residual gas, respectively, is
$0.2 \simless \epsilon$ $\simless 0.4$ \cite{LadaLada2003} implying that
the physics dominating the star-formation process on scales $<10$~pc
is stellar feedback.  Within this volume, the pre-cluster cloud core
contracts under self gravity thereby forming stars ever more
vigorously, until feedback energy suffices to halt the process ({\it
feedback-termination}), \cite{WeidnerKroupa2006}.  This occurs on one
to a few crossing times ($\approx 10^6$~yr), and since each proto-star
needs about $10^5$~yr to accumulate about 95~\% of its mass
\cite{WuchterlTscharnuter2003}, the assumption may be made that the
very young cluster is mostly virialised at feedback-termination.  Its
stellar velocity dispersion, $\sigma \approx \sqrt{G\,M_{\rm
ecl}/(\epsilon \, R)}$, may then reach $\sigma=40\,$pc/Myr if $M_{\rm
ecl} = 10^{5.5}\,M_\odot$ which is the case for $\epsilon\,R <
1$~pc. This is easily achieved since the radius of one-Myr old
clusters is $R\approx 1$~pc with a weak, if any dependence on mass.

The above exercise demonstrates that the possibility may be given that
a {\it hot kinematical component} could add to a galactic disk as a
result of clustered star formation for reasonable physical
parameters. But this depends on (i) $\epsilon$, (ii) $R$ (cluster
concentration) and (iii) the ratio of the gas-expulsion time-scale to
the dynamical time of the embedded cluster, $\tau_{\rm gas}/t_{\rm
cross}$.

\subsection{Empirical constraints}
\label{pk_ssec:emp}
The first (i) of these is clearly fulfilled: $\epsilon<40$~\%
\cite{LadaLada2003}.  The second (ii) also appears to be fulfilled
such that clusters with ages $\simless 1$~Myr have $R\simless1$~pc
independently of their mass. Some well-studied cases are tabulated and
discussed in \cite{Kroupa2005}. Finally, the ratio $\tau_{\rm
gas}/t_{\rm cross}$ (iii) remains uncertain but critical.

The well-observed cases discussed in \cite{Kroupa2005} do indicate
that the removal of most of the residual gas does occur within a
cluster-dynamical time, $\tau_{\rm gas}/t_{\rm cross} \simless
1$. Examples noted are the Orion Nebula Cluster (ONC) and R136 in the
LMC both having significant super-virial velocity dispersions. Other
examples are the Treasure-Chest cluster and the very young
star-bursting clusters in the massively-interacting Antennae galaxy
which appear to have HII regions expanding at velocities such that the
cluster volume may be evacuated within a cluster dynamical time.

A simple calculation of the amount of energy deposited by an O~star
into its surrounding cluster-nebula also suggests it to be larger than
the nebula binding energy \cite{Kroupa2005}. Furthermore,
\cite{BastianGoodwin2006} note that many young clusters have a
radial-density profile signature expected if they are expanding
rapidly.

Thus, the data suggest the ratio $\tau_{\rm gas}/t_{\rm cross}$ to be
near one, but much more observational work needs to be done to
constrain this number. Measuring the kinematics in very young clusters
would be an extremely important undertaking, because the implications
of $\tau_{\rm gas}/t_{\rm cross}\simless 1$ are dramatic.

To demonstrate these implications it is now assumed that a cluster is
born in a very compact state ($R\approx 1$~pc), with a low sfe
($\epsilon <0.4$) and $\tau_{\rm gas}/t_{\rm cross}\simless 1$. As
noted in \cite{Kroupa2005}, ``in the presence of O~stars, explosive
gas expulsion may drive early cluster evolution independently of
cluster mass''.

\section{Implications}
\label{pk_sec:impl}

\subsubsection{Heating galactic-field populations}
\label{pk_ssec:heat}
As one of the important implications, a cluster in the age range of
$\approx 1-50$~Myr will have an unphysical $M/L$ ratio because it is
out of dynamical equilibrium rather than having an abnormal stellar
IMF \cite{BastianGoodwin2006}. Another implication would be that a
Pleiades-like open cluster would have been born in a very dense
ONC-type configuration and that, as it evolves, a ``moving-group-I''
is established during the first few dozen~Myr which comprises roughly
2/3rds of the initial stellar population and is expanding outwards
with a velocity dispersion which is a function of the
pre-gas-expulsion configuration \cite{Kroupaetal2001}. These
computations were in fact the first to demonstrate, using
high-precision $N$-body modelling, that the re-distribution of energy
within the cluster during the embedded phase and the expansion phase
leads to the formation of a substantial remnant cluster despite the
inclusion of all physical effects that are disadvantageous for this to
happen (explosive gas expulsion, Galactic tidal field and mass loss
from stellar evolution).  A ``moving-group-II'' establishes later as
the ``classical'' moving group made-up of stars which slowly
diffuse/evaporate out of the re-virialised cluster remnant with
relative kinetic energy close to zero.

Thus, the moving-group-I would be populated by stars that carry the
initial kinematical state of the birth configuration into the field of
a galaxy.  Each generation of star clusters would, according to this
picture, produce overlapping moving-groups-I (and~II), and the overall
velocity dispersion of the new field population can be estimated by
adding in quadrature all expanding populations. This involves an
integral over the embedded-cluster mass function, $\xi_{\rm
ecl}(M_{\rm ecl})$, which describes the distribution of the stellar
mass content of clusters when they are born \cite{Kroupa2002,
Kroupa2005}.  Because the embedded cluster mass function is known to
be a power-law this integral can be calculated for a first estimate.
The result is that for reasonable upper cluster mass limits in the
integral, $M_{\rm ecl}\simless10^5\,M_\odot$, the observed
age--velocity dispersion relation of Galactic field stars can be
re-produced.

This theory can thus explain the much debated ``energy deficit'': that
the observed kinematical heating of field stars with age cannot, until
now, be explained by the diffusion of orbits in the Galactic disk as a
result of scattering on molecular clouds, spiral arms and the bar
\cite{Jenkins1992}. Because the age--velocity-dispersion relation for
Galactic field stars increases with stellar age, this notion can also
be used to map the star-formation history of the Milky-Way disk by
resorting to the observed correlation between the star-formation rate
in a galaxy and the maximum star-cluster mass born in the population
of young clusters \cite{Weidneretal2004}.

\subsubsection{Structuring the initial cluster mass function}
\label{pk_ssec:mfn}
Another potentially important implication from this theory of the
evolution of young clusters is that {\it if} the gas-expulsion
time-scale and/or the sfe varies with initial (embedded) cluster mass,
then an initially featureless power-law mass function of embedded
clusters will rapidly evolve to one with peaks, dips and turnovers at
``final'' cluster masses that characterize changes in the broad
physics involved, such as the gas-evacuation time-scale.

As an example, \cite{KroupaBoily2002} assumed that the function
$M_{\rm icl} = f_{\rm st}\,M_{\rm ecl}$ exists, where $M_{\rm ecl}$ is
as above, $M_{\rm icl}$ is the ``classical initial cluster mass'' and
$f_{\rm st} = f_{\rm st}(M_{\rm ecl})$. The ``classical initial
cluster mass'' is that mass which is inferred by classical $N$-body
computations without gas expulsion (i.e. in effect assuming
$\epsilon=1$, which is however, unphysical). Thus, for example, for
the Pleiades, $M_{\rm cl}\approx 1000\,M_\odot$ at the present time
(age: about 100~Myr), and a classical initial model would place the
initial cluster mass to be $M_{\rm icl}\approx 1500\,M_\odot$ by using
standard $N$-body calculations to quantify the secular evaporation of
stars from an initially bound and virialised ``classical'' cluster
\cite{Portetal2001}. If, however, the sfe was 33~per cent and the
gas-expulsion time-scale was comparable or shorter than the cluster
dynamical time, then the Pleiades would have been born in a compact
configuration resembling the ONC and with a mass of embedded stars of
$M_{\rm ecl}\approx 4000\,M_\odot$ \cite{Kroupaetal2001}.  Thus,
$f_{\rm st}(4000\,M_\odot) = 0.38$.

By postulating that there exist three basic types of embedded
clusters, namely clusters without O~stars (type~I: $M_{\rm
ecl}\simless 10^{2.5}\,M_\odot$, e.g.  Taurus-Auriga pre-main sequence
stellar groups, $\rho$~Oph), clusters with a few O~stars (type~II:
$10^{2.5} \simless M_{\rm ecl}/M_\odot \simless 10^{5.5}$, e.g. the
ONC) and clusters with many O~stars and with a velocity dispersion
comparable to the sound velocity of ionized gas (type~III: $M_{\rm
ecl}\simgreat 10^{5.5}\,M_\odot$) it can be argued that $f_{\rm
st}\approx 0.5$ for type~I, $f_{\rm st}<0.5$ for type~II and $f_{\rm
st}\approx 0.5$ for type~III. The reason for the high $f_{\rm st}$
values for types~I and~III is that gas expulsion from these clusters
may be longer than the cluster dynamical time because there is no
sufficient ionizing radiation for type~I clusters, or the potential
well is too deep for the ionized gas to leave (type~III
clusters). Type~II clusters undergo a disruptive evolution and witness
a high ``infant mortality rate'' \cite{LadaLada2003}, therewith being
the pre-cursors of OB associations and open Galactic clusters.

Under these conditions and an assumed functional form for $f_{\rm
st}=f_{\rm st}(M_{\rm ecl})$, the power-law embedded cluster mass
function transforms into a cluster mass function with a turnover near
$10^5\,M_\odot$ and a sharp peak near $10^3\,M_\odot$
\cite{KroupaBoily2002}. This form is strongly reminiscent of the
initial globular cluster mass function which is inferred by
e.g. \cite{Vesperini1998,Vesperini2001,ParmentierGilmore2005,Baumgardt2006}
to be required for a match with the evolved cluster mass function that
is seen to have a universal turnover near $10^5\,M_\odot$.

This ansatz may thus bear the solution to the long-standing problem
that the deduced initial cluster mass function needs to have this
turnover, while the observed mass functions of young clusters are
feature-less power-law distributions.

%
%
%
%

%
%



\printindex
\end{document}